\documentstyle[12pt]{article}

\def\d0{\Delta^0}
\def\dm{\Delta^{++}}
\def\tt{\frac{3}{2}\frac{3}{2}}
\def\wt{\widetilde}
\def\be{\begin{equation}}
\def\ee{\end{equation}}
\def\mm{{\rm ++}}
\begin{document}
\begin{flushright}
\large{CINVESTAV-FIS-14/95}
\end{flushright}
\vspace{0.6cm}
\begin{center}
\LARGE{\bf Pion-proton scattering and isospin breaking in the  $\d0-\dm$
system}
 \end{center}
\vspace{.8cm}
\begin{center}
\Large A. Bernicha$^1$, G. L\'opez Castro$^2$ and J. Pestieau$^1$\\
\vspace*{.4cm}
{\normalsize{\it $^1$ Institut de Physique Th\'eorique, Universit\'e
 Catholique \\ \it de Louvain, B-1348 Louvain-la-Neuve, BELGIUM. \\
\it $^2$ Departamento de F\'\i sica, Cinvestav del IPN, Apdo. \\
\vspace{-.4cm} \it  Postal 14-740, 07000 M\'exico, D.F., MEXICO.}}
\vspace*{.4cm}
\end{center}

\thispagestyle{empty}
\vspace{.4cm}
\centerline{ \bf Abstract}
\vspace{.3cm}
 We determine the mass and width of the  $\Delta^{++}\
(\Delta^0)$ resonance from data on $\pi^+ p\ (\pi^- p)$ scattering both,
in the pole of the $S$-matrix and conventional Breit-Wigner approaches to
the scattering amplitude. We provide a simple formula
 that  relates the two definitions for the parameters of the $\Delta$.
 Isospin symmetry breaking in the $\d0-\dm$ system depends on the
definition of the resonant properties: we find $M_0-M_{++} = 0.40 \pm
0.57\ {\rm MeV},\ \Gamma_0 -\Gamma_{++} = 6.89 \pm 0.95\ {\rm MeV}$ in
the pole approach while $\wt{M}_0-\wt{M}_{++} = 2.25 \pm 0.68\ {\rm
MeV},\ \wt{\Gamma}_0 - \wt{\Gamma}_{++} = 8.45 \pm 1.11\ {\rm MeV}$ in
the conventional approach.

\vspace{.5cm}
\vspace{1cm}

\newpage
\setcounter{page}{1}
\vspace{2cm}

\begin{center}
\bf I. Introduction.
\end{center}

 The isospin symmetry of strong interactions is a very good approximation
to relate  some properties and processes involving hadrons of a given
isospin multiplet. The reason for this is that, at the fundamental level,
the isospin symmetry is broken only by the electromagnetic interactions
and the mass difference of the $u$ and $d$ quarks. However, it is not
easy to perform a precise theoretical calculation for isospin breaking
effects in hadrons starting from the fundamental theory; for instance, the
old problem of the neutron-proton mass difference [1] (which has been
measured with an accuracy of 7 parts per million [2]) remain as a challenge
for the theory of elementary particles.

  In this work we are concerned with the isospin breaking in the masses
and widths of the $\d0,\ \dm$ members of the $I=J=3/2$ multiplet of
baryon resonances. As is well known, these resonances would have equal
masses and widths if isospin symmetry were exact.
 Actually, the $\Delta$'s undergo strong
interaction decays to $N\pi$ final states with branching fractions larger
than 99 \% [2].

  Experimentally, the tests of isospin symmetry in the $\Delta$ system
faces the problem that the definition of mass and width for an unstable
particle is not unique. In fact, there are two common approaches to
extract these resonance parameters from experimental data. In the {\em
conventional} approach, the transition amplitude is parametrized in terms
of a Breit-Wigner containing an energy-dependent width. A partial wave
analysis of this amplitude allow to define the mass $\wt{M}$ as the energy
where the phase shift attains 90$^0$. From this, the width is defined as
$\wt{\Gamma}(E=\wt{M})$.
   On the other side, the {\em pole} approach allows to define the mass
$M$ and width $\Gamma$ of the resonance from the real and imaginary parts
of the pole position in the $S$-matrix amplitude.

   The pole position is believed to be a physical property of the
$S$-matrix amplitude [3, 4] and to provide a definition for the mass and
width of a resonance which is independent of the physical process used to
extract these parameters. In contrast, in the conventional
approach one requires to model the production and decay of the
resonance {\em i.e.}, the energy dependence of the decay width involved in
the amplitude.

   In this paper we use the experimental data on $\pi^{\pm} p$ scattering
[5] to extract the pole parameters of the $\d0,\ \dm$ baryon resonances.
It is found that the isospin splittings in the $\d0-\dm$ system is
different for both definitions of the resonant parameters: the resonant
parameters in the conventional approach exhibit a stronger isospin
breaking that in the $S$-matrix pole approach. Also, a simple formula is
provided to relate the resonant parameters defined in the two approaches.

   The remaining of this paper is organized as follows. In section II we
describe the two approaches for the $\pi p$ scattering amplitude in the
$\Delta$ resonance region. In sections III and IV we analyse,
respectively, the $\pi^+p$ and $\pi^-p$ scattering in order to  extract the
resonant parameters of the $\dm$ and $\d0$. Section V contains a
discussion of our results and conclusions and an Appendix is devoted to
repeat the analysis of sections III and IV in the case of
`non-relativistic' pole scattering amplitudes.

\

\begin{center}
{\bf II. Pole and conventional approaches to the $\Delta$ in $\pi p$
scattering}
\end{center}

\

 In this section we discuss in more detail the two approaches for the
description of the $\Delta$ resonance in $\pi p$ scattering. We also
derive the relations to pass from the resonance parameters in one
approach to the other.

  The total cross section for $\pi p$ scattering in the $\Delta$
resonance region can be written in terms of the partial wave amplitude
$a_{\tt}$ as [see for example, p.1293 in Ref. 2]:
\be
\sigma_{\tt}(\pi p) = \frac{8\pi}{k^2}\ |a_{\tt}|^2
\ee
where $k$ denotes the center of mass momentum of either $\pi$ or $p$.

   For elastic scattering, the partial wave amplitude can be written in
terms of the corresponding phase shift $\delta_{\tt}$:
\be
a_{\tt} = \frac{\tan \delta_{\tt}}{1-i\tan \delta_{\tt}}
\ee
which automatically satisfies unitarity.

  In the conventional approach, the amplitude $a_{\tt}$ is saturated with
the $\Delta$ resonance which is described by an energy dependent width
$\wt{\Gamma}(s)$, where $s$ denotes the squared center of mass energy. If
the phase shift is chosen as
\be
\tan \delta_{\tt} = -\ \frac{\wt{M} \wt{\Gamma}(s)}{s-\wt{M}^2},
\ee
we are lead to the usual Breit-Wigner form of the amplitude, namely:
\be
a_{\tt}=-\ \frac{\wt{M} \wt{\Gamma}(s)}{s-\wt{M}^2+i\wt{M}\wt{\Gamma}(s)}.
\ee
Thus, the mass and width of the $\Delta$ in the conventional approach
become, respectively, $\wt{M}$ and $\wt{\Gamma}(s=\wt{M}^2)$.

  As it was mentionned above, the $S$-matrix approach provides a
definition for the parameters of an unstable particle which is
independent of the process used to extract them. This happens
because, independently of the specific scattering or decay process, the
resonance
shows up in the amplitude as a physical pole $\overline{s}$. In this
approach, the resonant and background contributions (in the {\em same}
channel) to the amplitude are explicitly separated according to [3]:
\be
a=\frac{R}{s-\overline{s}} + B.
\ee
Thus, the
mass $M$ and width $\Gamma$ of the resonance in the pole approach are
defined as follows [3] (see also the Appendix and Ref. [6] for an
alternative definition):
\be
\overline{s} \equiv M^2 - iM\Gamma.
\ee

  In order to connect the two approaches, let us split the phase shift
$\delta_{\tt}$ into two terms:
\be
\delta_{\tt} = \delta_R + \delta_B
\ee
where $\delta_R$ corresponds to the phase shift due to the $\Delta$
resonance and $\delta_B$ to the background contribution in the
($\frac{3}{2}, \frac{3}{2}$)
channel. The choice in Eq. (7), explicitly leads to the scattering
amplitude of the form given in Eq. (5) (see Ref. [7] and Eq. (12) below).

  Since the background is expected to give a small contribution to the
$\pi p$ scattering amplitude in the resonance region, we can choose the
following parametrization [7, 8]:
\be
\tan \delta_B = x(s)
\ee
where $x(s)$ represents a smooth function of $s$.

  If we define
\be
\tan \delta_R = -M\Gamma/(s-M^2)
\ee
for the resonance
contribution and if we introduce Eqs. (7)-(9) into Eq. (2) we are lead to
the following equivalent representations for the amplitude:
\begin{eqnarray}
a_{\tt} &=& -\ \frac{M\Gamma - x(s)(s-M^2)}{[1-ix(s)](s-M^2+iM\Gamma)} \\
\vspace{.5cm} &=& -\ \frac{[M\Gamma-x(s)(s-M^2)]}{s-M^2+x(s)M\Gamma
+i[M\Gamma -x(s)(s-M^2)]} \\
\vspace{.5cm} &=& -\ \frac{M\Gamma}{s-M^2+iM\Gamma}\exp(2i\delta_B) +
 \frac{x(s)}{1-ix(s)}. \end{eqnarray}

  If we compare Eqs. (11) and (2) we immediatly get the identity:
\begin{eqnarray}
\tan \delta_{\tt} &=& -\ \frac{M\Gamma - x(s)(s-M^2)}{s-M^2+x(s)M\Gamma} \\
   &=&-\ \frac{\wt{M} \wt{\Gamma}(s)}{s-\wt{M}^2}.
\end{eqnarray}

  Since the resonant parameters in the conventional approach are defined
according to $\delta_{\tt}=90^0$ when $s=\wt{M}^2$, from the previous
equations we obtain the relations between the resonant parameters in
both approaches, namely [7]:
\be
\wt{M}^2 = M^2-xM\Gamma
\ee
and
\be
\wt{M}\wt{\Gamma}=M\Gamma(1+x^2)/(1+M\Gamma x')
\ee
where $x,\ \wt{\Gamma}$ and $x'=dx/ds$ are evaluated at $s=\wt{M}^2$. Eqs.
(15)-(16)
will allow us to extract $\wt{M}$ and $\wt{\Gamma}$ from the fitted
values of $M,\ \Gamma$ and $x$ (see sections III and IV).

 Defining
\[
M\Gamma(s) \equiv M\Gamma - x(s) (s-M^2),
\]
we get
\be
x(s) = -\ M\Gamma\ \left (\frac{\gamma(s)-1}{s-M^2} \right)
\ee
where
\be
\gamma(s)=\Gamma(s)/\Gamma
\ee
with $\gamma(M^2)=1$. The $s$-dependence of the total width
$\Gamma(s)$ (or equivalently $x(s)$) will be
introduced later\footnote{We would like to emphasize that various
parametrizations for $x(s)$ were used to fit the $\pi p$ experimental
data (for instance, we used the parametrizations of Ref. [9] for the
background contributions to $e^+e^- \rightarrow \pi^+ \pi^-$). As
expected, these background parametrizations {\em do not} modify the
position of the pole.}
 .
Note that $x(s)$ is a regular function when $s$
approaches $M^2$.

  With the above choice for $x(s)$, Eq. (10) becomes:
\begin{eqnarray}
a_{\tt} &=&  -\ \frac{M\Gamma(s)}{[1-ix(s)](s-M^2+iM\Gamma ]} \\
\vspace{.5cm} &=& -\ \frac{M\Gamma(s)}{s-M^2+x(s)M\Gamma + iM\Gamma(s)}
\end{eqnarray}
which looks very similar to the usual Breit-Wigner parametrization, Eq.
(4),  if we define an effective
mass $M_{eff}^2\approx M^2-xM\Gamma$ because $x(s)$ varies smoothly
around the resonance.

\

\begin{center}
\bf III. Analysis of the $\pi^+ p$ scattering.
\end{center}

  In this section we perform the fit of the experimental data on $\pi^+
p$ scattering [5] to extract the $\dm$ parameters by using the formalism
described in the previous section.

   The total cross section for $\pi^+ p$ scattering in the $(I,J) =
(\frac{3}{2}, \frac{3}{2})$ channel is given by:
\be
\sigma_{\tt}(\pi^+p) = \frac{8\pi}{k^2}\ | a_{\tt}^{\mm}|^2.
\ee
 As discussed in section II, the scattering amplitude $a_{\tt}^{\mm}$ can
be written as:
\be
a_{\tt}^{\mm}=-\
\frac{M_{\mm}\Gamma_{\mm}-x_{\mm}(s)(s-M_{\mm}^2)}{[1-ix_{\mm}(s)](s-
M_{\mm}^2+iM_{\mm}\Gamma_{\mm})}
\ee
where $x_{++}(s)$ is given by:
\be
x_{\mm}(s)=-\ M_{++}\Gamma_{++}\left( \frac{\gamma_{++}(s)-1}{s-M_{++}^2}
\right). \ee

   We choose $\gamma_{++}(s)$ to be the standard parametrization for
the energy-dependent width used in the experiments as given, for example,
in Ref. [5]:
\be
\gamma_{++}(s)=\left( \frac{k}{k_{++}} \right)^3\,
\frac{1+a_{++}(k_{++}/m_{\pi^+})^2}{1+a_{++}(k/m_{\pi^+})^2}
\ee
where $k$ denotes the center of mass momentum of $\pi^+$ and $k_{++}$ the
value of $k$ at $\sqrt{s}=M_{++}$. $a_{++}$ is a dimensionless parameter.

  Thus, Eq. (22) contains three free parameters to be adjusted from the
$\pi^+p$ experimental data: the pole resonance parameters $(M_{++},\
\Gamma_{++})$ and the parameter $a_{++}$. The fitted values for these
quantities allow to extract the resonance parameters in the conventional
approach by using Eqs. (15) and (16).

  In the fit to the experimental data of Ref. [5] we distinguish two cases:
\vspace{-1.0cm}
\begin{quote}
\item (A) We first take into account the background contributions coming
from
channels {\em other} than $(I,\ J)  = (\frac{3}{2},\ \frac{3}{2})$ as
given in the last column-Table 1 of Ref. [5].
\item(B) The same as before but we allow a 10 \% error for the background
contributions of Table 1 in Ref. [5].
\end{quote}
 The results of the fits are shown in Table 1 and the fit for case (A)
is also shown in Fig. 1. The following remarks are in order:
\begin{enumerate}
\item The mass and width of the $\dm$ in the pole approach are shifted to
lower values by around 20 and 12 MeV, respectively, with respect to the
resonant parameters in the conventional approach.
\item With the parameters shown in Table 1 and using Eqs. (23)-(24), we
can easily check that the variation of $x_{++}(s)$ in the kinematical region
$1100\ {\rm MeV} < \sqrt{s} < 1300\ {\rm MeV}$ is less than 10 \%.
\item The most important effect of considering case (B) is observed in
the parameter $a_{++}$.
\end{enumerate}

  The pole parameters $M_{++},\ \Gamma_{++}$ shown in Table 1 are a
little bit different from other available determinations which are shown
in Table 2 (our results are repeated for comparison).

  Similarly, the mass and width values shown in Table 1 for the $\dm$ in the
conventional approach are very similar to the following results of Ref.
[5]:
\begin{eqnarray}
\wt{M}_{++} &=& 1232.1 \pm 0.2\ {\rm MeV} \nonumber \\
 \wt{\Gamma}_{++} &=& 109.8 \pm 0.4\ {\rm MeV}.
\end{eqnarray}

\

\begin{center}
\bf IV. The $\d0$ in $\pi^-p$ scattering.
\end{center}

 In this section we apply the formalism described in section II to the
production of the $\d0$ in $\pi^- p$ scattering.
  The analysis of $\pi^-p$ scattering is slightly more
complicated because both, $\pi^-p$ and $\pi^0n$, can be reached as final
states. Thus, due care of isospin breaking coming from the
$\pi^+-\pi^0$ and $n-p$ mass differences and possible residual isospin
breaking effects have to be taken into account.

  As in the previous case, the total cross section for $\pi^-p$
scattering in the $(I,\ J)=(\frac{3}{2},\ \frac{3}{2})$ channel is given by:
\be
\sigma_{\tt}(\pi^-p) = \frac{8\pi}{k^2}\ |a_{\tt}^0|^2.
\ee

   In order to incorporate isospin breaking effects we first realize that
in the limit of isospin symmetry we would have:
\be
|a_{\tt}^0|^2 = \frac{1}{3}\ |a_{\tt}^{++}|^2
\ee
and also $M_{\d0}=M_{\dm},\ \Gamma_{\d0}=\Gamma_{\dm}$ (the superindex in
$a$  refers to the charge of the $\Delta$, and $M$ and $\Gamma$ are the
resonant parameters of the $\Delta$).
 Observe that, apart from the small
radiative decay $\d0 \rightarrow n\gamma$ ($BR(\d0 \rightarrow n\gamma)
\sim 0.55\ {\rm to}\ 0.61$ \% [2]), the $\Delta$'s undergo strong
interaction decays to $\pi N$.

  The isospin breaking can be taken into account by properly modifying
Eq. (19), namely by using:
\be
|a_{\tt}^0 |^2 = \frac{M_0^2 \Gamma_{\d0 \rightarrow p\pi^-}(s)
\Gamma_0(s)}{(1+x_0^2(s)) |s-M_0^2 +iM_0\Gamma_0|^2}
\ee
where,
\be
\Gamma_0(s) = \Gamma_{\d0 \rightarrow p\pi^-}(s) + \Gamma_{\d0
\rightarrow n\pi^0}(s)
\ee
when we neglect the tiny $\d0 \rightarrow n\gamma$ contribution to the
total width of $\d0$.

   The partial decay widths of the $\d0$ can be written as follows:
\begin{eqnarray}
\Gamma_{\d0 \rightarrow p\pi^-}(s) &=& \frac{1}{3}(1+\epsilon)\Gamma_0
\gamma_-(s) \\     \vspace{.5cm}
\Gamma_{\d0 \rightarrow n\pi^0}(s) &=& \frac{2}{3}(1-\frac{\epsilon}{2})
\Gamma_0 \gamma_0(s)
\end{eqnarray}
where ($i=-,\ 0$),
\be
\gamma_i(s) = \left( \frac{k_i}{k_i^0} \right)^3 \ \frac{1+a_0
(k_i^0/m_{\pi^+})^2}{1+a_0(k_i/m_{\pi^+})^2}.
\ee
$k$ ($k^0)$ denotes the center of mass momentum of either one of the final
particles coming
from the $\d0$ at$\sqrt{s}$  ($\sqrt{s}=M_0$). The small
dimensionless parameter $\epsilon$ takes into account possible residual
effects of isospin breaking.

   If we neglect second order isospin breaking effects of ${\cal O}(
\epsilon [\gamma_-(s)-\gamma_0(s)])$ in Eq. (29), we obtain the following
expression for the total width:
\be
\Gamma_0(s) \approx \Gamma_0 \left\{ \frac{1}{3}\gamma_-(s) +
\frac{2}{3}\gamma_0(s)\right\}
\ee
and the expression for the background contribution becomes:
\be
x_0(s) = -\ \frac{M_0 \Gamma_0}{s-M_0^2}\ \left( \frac{1}{3} \gamma_-(s)
+\frac{2}{3}\gamma_0(s) -1 \right) .
\ee

  The set of four free parameters ($M_0,\ \Gamma_0,\ a_0,\ \epsilon$) can be
determined from a fit to the $\pi^-p$ experimental data of Ref. [5] by
using Eqs. (26) and (28--34).
\newpage
 As for $\pi^+p$ scattering, we
have considered two cases in the fit:
\vspace{-1.0cm}
\begin{quote}
\item (C) We have used the experimental data on the $\pi^- p$ cross section
and the
background contributions coming from channels other than ($\frac{3}{2},\
\frac{3}{2}$) as given in Table 1 of Ref. [5].
\item (D) The same as before but we attribute a $\pm 10 \%$ error to the
background contributions.
\end{quote}

  The results of the fits are shown in Table 3 and Fig. 2 (case C). From
Table 3 we can draw the following conclusions:
  \begin{enumerate}
\item The position of the pole remains the same for cases (C) and (D).
 The most important effect of attributing a 10 \% error to the
background contributions is observed in the dimensionless
parameter $a_0$ appearing in the expression for $x_0(s)$.
\item The mass and width of the $\d0$ in the pole approach are shifted to
lower values for about 22 and 14  MeV, respectively, respect to the
values of those parameters in the conventional approach.
\item The residual isospin breaking parameter $\epsilon$ is of the
expected order of magnitude.
\end{enumerate}

   The values of the mass and width pole parameters ($M_0,\ \Gamma_0$) can
be compared with other determinations of these resonant properties of the
$\d0$ as shown in the Table 4.

   The values of our fit for the $\d0$ parameters in the conventional
approach (see Table 3) as derived from Eqs. (15) and (16) are very
similar to the corresponding parameters of Pedroni {\em et. al.} [5]:
\begin{eqnarray}
\wt{M}_0 &=& 1233.5 \pm 0.2\ {\rm MeV} \\
\wt{\Gamma}_0 &=& 118.4 \pm 0.9\ {\rm MeV}
\end{eqnarray}

\

\begin{center}
\bf V. Discussion of results and conclusions.
\end{center}

   In this work we have analysed the experimental data on $\pi p$
scattering [5] in the $\Delta$ resonance region in order to get
information about the isospin breaking in the resonant parameters of the
$\dm$ and $\d0$. For these purposes, we have explicitly separated, in the
($\frac{3}{2}, \frac{3}{2}$) channel, the
pole and background contributions to the scattering amplitudes and we have
obtained simple expressions that relate the resonant properties of the
$\Delta$'s in the pole and conventional approaches.

   Our main results are summarized in Tables 1 and 3. Our results for the
pole parameters of the $\dm$ and $\d0$ are independent of the
precise choice to
parametrize the background contribution through the smooth function
$x(s)$, as it should be.
 From these tables we can obtain the isospin breaking in the masses and
widths of the $\Delta$'s and we compare our results with other available
determinations of these quantities in Table 5 (all entries  are given in
MeV).

  In our analysis we have considered the background contributions as
given in Table 1 of Ref. [5] (case I) and we have repeated the analysis
by adding a $\pm 10 \%$ error to these backgrounds (case II).

   Regarding the isospin breaking in the $\d0 - \dm$ system we conclude
the following from Table 5:

1). The isospin breaking is larger in the resonant parameters defined in
the conventional approach than in the pole approach.  The available
determinations of isospin breaking in the masses in either of the
approaches are rather similar while isospin breaking in the widths spreads
over a wider range.

2). The isospin breaking in the pole masses of the $\Delta$'s is
consistent with zero ($M_0 \approx M_{++}$). This result differs from the
naive expectation based on rough estimates of mass difference coming from
 electromagnetic and
$m_d-m_u$ contributions. Indeed, using the expression for the
neutron--proton mass difference \be
m_n-m_p = (\delta m)_{em} + c(m_d-m_u)
\ee
and since the quark content of $\dm$ and $\d0$  are $uuu$ and $ddu$,
respectively, we would roughly expect
\begin{eqnarray}
M_{\d0}-M_{\dm} & = & 2(\delta m)_{em} + 2c(m_d-m_u) \nonumber \\
 & = & 2 (m_n-m_p) \nonumber \\
 & \approx & 2.6\ {\rm MeV}.
\end{eqnarray}
  In contrast, the isospin breaking in the masses of the $\Delta$'s
defined in the conventional approach are in agreement with the naive
expectation of Eq. (38). Note that the pole mass is the correct way to
define a physical mass [3].

  3). From Table 5 we observe that our results exhibit an isospin
breaking of about 7 \% in the total widths of the $\Delta$'s.

This isospin
breaking can also be measured through the background contribution at
threshold ($s_{th}=(m_p + m_{\pi^{\pm}})^2 \approx (m_n + m_{\pi^0})^2$).
More explicitly, since
$M_0 \approx M_{++}$ and $\gamma(s_{th})=0$ it follows from Eq. (17) and
(34) that
\[
x(s_{th})=M\Gamma/(s_{th}-M^2)
\]
or
\be
\frac{x_0(s_{th})}{x_{++}(s_{th})}\approx \frac{\Gamma_0}{\Gamma_{++}}
\approx 1.07
\ee
for the ratio of background contributions.

The numerical value in Eq. (39)  follows from $x_0(s_{th})
= -0.4084$ and $x_{++}(s_{th})=-0.3828$, which are obtained using the
results of Tables 1 (case A) and 3 (case C), respectively.
   As we have pointed out in the text, $x(s)$ is a slowly varying
function around the resonance. However it is interesting to observe that
it clearly reflects the breaking of isospin symmetry at threshold. As a
comparison, let us mention that the corresponding ratio at
$s=\wt{M}^2$ gives $x_0/x_{++}\approx 1.02$, which
exhibits a smaller isospin breaking. Tables 1 and 3 show that isospin
symmetry  breaking is much smaller in the parameters $x_i(\wt{M}^2)$ than
in the $a_i$'s. Our results are rather insensitive to the exact values of
the $a_i$'s.

  4). As written above, we have neglected the decay $\d0 \rightarrow n +
\gamma$. We have verified that, since $BR(\d0 \rightarrow n + \gamma )
\leq 0.6 \%$ [2], to neglect this mode does not affect our results because
isospin symmetry breaking in the total widths of the $\Delta$'s amounts
for 7 \%.

\

\appendix

\begin{center}
\bf Appendix.
\end{center}
  In this appendix we repeat the analysis of sections III and IV for the
case of a `non-relativistic' definition of the pole parameters. As can be
concluded by comparing Tables (6) and (7) with Tables (1) and (3), the main
conclusions of this paper are not modified by this assumption.

  As is well known [6], an alternative definition for the parameters of an
unstable particle in
the S-matrix approach is obtained by assuming that the phase shift
associated to the resonance is given by:
\be
\tan \delta_R =-\ \frac{\Gamma/2}{\sqrt{s}-M}.
\ee
  As it will become explicit later (see Eq. (43)), Eq. (40) gives
rise to an amplitude with the pole position at
\be
\sqrt{\overline{s}}=M-\frac{i}{2}\Gamma.
\ee
Eq. (40) can be obtained from Eq. (9) by replacing
\be
s-M^2 \rightarrow 2M(\sqrt{s}-M).
\ee
Note that $s-M^2 \approx 2M(\sqrt{s} - M)$ is a good approximation for
values of $\sqrt{s}$
close to the resonance. This is the reason for calling Eq. (41) a
non-relativistic definition of the pole parameters.

   With the above definition for $\delta_R$, the analogous of Eqs. (22) and
(28) become, respectively:
\be
a_{\tt}^{++} = -\
\frac{\Gamma_{++}/2-x_{++}(s)(\sqrt{s}-M)}{[1-ix_{++}(s)](\sqrt{s}-M_{++}
+ i\Gamma_{++}/2)}
\ee
and
\be
|a_{\tt}^0|^2 = \frac{1}{4}\cdot \frac{\Gamma_{\d0 \rightarrow p\pi^-}(s)
\Gamma_0(s)}{(1+x_0^2(s))|\sqrt{s}-M_0 + i\Gamma_0/2|^2}.
\ee
  The relations -- Eqs. (15) and (16) -- between the resonant
parameters in both approaches are also modified to become:
\begin{eqnarray}
\wt{M} &=& M-x\Gamma/2 \\
\wt{\Gamma} &=& \Gamma (1+x^2)/(1+\frac{\Gamma}{2} x')
\end{eqnarray}
where, $x,\ \wt{\Gamma}$ and $x'=dx/d\sqrt{s}$, are evaluated at
$s=\wt{M}^2$.

  In order to fit the experimental data of Ref. [5] we have, as in
sections III and IV, distinguished two cases: $(a)$ we use the data on
$\pi^+ p$ and $\pi^- p$ scattering by considering also the background
contributions as given in Table 1 of Ref. [5] and, $(b)$ the same as
before but we attribute a $\pm 10\ \%$ error to the background.

   The results of the fits are shown in Table 6 for the $\dm$ and in
Table 7 for the $\d0$. We observe that the parameters of Tables 6 and 7
agree to a
high accuracy with the values in the relativistic definition shown in
Tables 1 and 3. In fact we observe that $M_{\Delta}({\rm
``relativistic"}) \approx M_{\Delta}({\rm ``non relativistic"})-\ {\rm 1\
MeV},\ \Gamma_{\Delta}({\rm ``relativistic}) \approx \Gamma_{\Delta}({\rm
``non relativistic"}) +\ {\rm 1\ MeV}$.

   From Tables (6) and (7), the isospin breaking in the pole parameters are:
\begin{eqnarray}
M_0 - M_{++} &=& 0.70 \pm 0.58 \ {\rm MeV} \\
\Gamma_0 - \Gamma_{++} &=& 6.81 \pm 0.91\ {\rm MeV}
\end{eqnarray}
for case $(a)$ and
\begin{eqnarray}
M_0 - M_{++} &=& 0.80 \pm 0.76 \ {\rm MeV} \\
\Gamma_0 - \Gamma_{++} &=& 6.51 \pm 1.02\ {\rm MeV}
\end{eqnarray}
for case $(b)$, which are very similar to the results shown in Table 5.

\

\begin{center}
\bf Note added
\end{center}

  After we have completed this work we became aware of reference [14],
where expressions that relate the resonance parameters in the pole and
conventional approaches are also provided for the $N$'s and $\Delta$'s
(see Eqs. (3) and (A2) in Ref. [14]). The values quoted for the pole
parameters of the {\em generic} $\Delta$ resonance using his Eqs. (3) and
(A2) are similar to ours. Isospin breaking is not considered in Ref.
[14].

\newpage
\begin{center}
TABLE CAPTIONS
\end{center}

\begin{enumerate}
\item Resonant parameters of the $\dm$ extracted from $\pi^+ p$ scattering
. The values of $\wt{M}_{++}$ and $\wt{\Gamma}_{++}$ are obtained
using Eqs. (15), (16).
\item Comparison of our results for the pole parameters of the $\dm$ with
other available determinations.
\item Resonant parameters of the $\d0$ extracted from $\pi^- p$ scattering
. The values of $\wt{M}_0$ and $\wt{\Gamma}_0$ are obtained
using Eqs. (15), (16).
\item Comparison of our results for the pole parameters of the $\d0$ with
other available determinations.
\item Isospin breaking in the mass and widths of the $\d0-\dm$ baryons. The
two cases (I and II) for our results are described in section V. All the
quantities are given in MeV units.
\item Resonant parameters of the $\dm$ extracted from $\pi^+ p$ scattering
. The values of $\wt{M}_{++}$ and $\wt{\Gamma}_{++}$ are obtained
using Eqs. (45), (46).
\item Resonant parameters of the $\d0$ extracted from $\pi^- p$ scattering
. The values of $\wt{M}_0$ and $\wt{\Gamma}_0$ are obtained
using Eqs. (45), (46).
  \end{enumerate}
\

\begin{center}
FIGURE CAPTIONS
\end{center}
\begin{enumerate}
\item Total cross section for the $\pi^+ p$ scattering as a funtion of
kinetic energy in the lab system. The solid line is
 our fit using the pole parameters given in Table 1 (case A).
\item Total cross section for the $\pi^- p$ scattering as a function of the
kinetic energy in the lab system. The solid line is
 our fit using the pole parameters given in Table 3 (case C).
\end{enumerate}
\newpage

\

\

\begin{table}[h]
\begin{tabular}{|c|c|c|}
\multicolumn{3}{c}{Table 1 } \\
\hline
 & Case A & Case B \\
\hline\hline
$M_{++}$ (MeV)& $1212.20 \pm 0.23$ & $1212.50 \pm 0.24$ \\
\hline
$\Gamma_{++}$ (MeV) & $97.06 \pm 0.35$ & $97.37 \pm 0.42$ \\
\hline
$a_{++}$ & $0.5978 \pm 0.0155$ & $0.6256 \pm 0.0203$ \\
\hline
$x_{++} (\wt{M}^2)$ & $- 0.4062 \pm 0.0015$ & $- 0.4012 \pm 0.0017$ \\
\hline
$\wt{M}_{++}$ (MeV) & $1231.75 \pm 0.27$ & $1231.88 \pm 0.29$ \\
\hline
$\wt{\Gamma}_{++}$ (MeV) &$109.85 \pm 0.41$ & $109.07 \pm 0.48$ \\
\hline
\end{tabular}
\end{table}

\

\

\begin{table}[h]
\begin{tabular}{|c|c|c|}
\multicolumn{3}{c}{Table 2 } \\
\hline
$M_{++}$ (MeV) & $\Gamma_{++}$ (MeV) & References \\
\hline\hline
$1210.9 \pm 0.8$ & $99.2 \pm 1.5$ & [10] \\
\hline
$1210.7 \pm 0.16$ & $99.21 \pm 0.23$ & [11] \\
\hline
$1209.6 \pm 0.5$ & $100.8 \pm  1.0$ & [12] \\
\hline
$1212.20 \pm 0.23$ & $97.06 \pm 0.35$ & our results case A \\
\hline
$1213.30 \pm 0.23$ & $96.17 \pm 0.34$ & our results case (a)\\
\hline
\end{tabular}
\end{table}
\newpage

\

\

\begin{table}[h]
\begin{tabular}{|c|c|c|}
\multicolumn{3}{c}{Table 3 } \\
\hline
 & Case C & Case D \\
\hline\hline
$M_0$ (MeV)& $1212.60 \pm 0.52$ & $1213.20 \pm 0.66$ \\
\hline
$\Gamma_0$ (MeV) & $103.95 \pm 0.88$ & $104.10 \pm 1.01$ \\
\hline
$a_0$ & $0.6914 \pm 0.0477$ & $0.7408 \pm 0.0611$ \\
\hline
$x_0 (\wt{M}^2)$ & $ - 0.4154 \pm 0.0035$ & $- 0.4099 \pm 0.0040$ \\
\hline
$\wt{M}_0$ (MeV) & $1234.00 \pm 0.62$ & $1234.35 \pm 0.75$ \\
\hline
$\wt{\Gamma}_0$ (MeV) &$ 118.30 \pm 1.03$ & $117.58 \pm 1.16$ \\
\hline
$\epsilon$ ($\times$ 10$^{-2}$) & $2.2 \pm 0.3$ & $2.5 \pm 0.4$ \\
\hline
\end{tabular}
\end{table}

\

\

\begin{table}[h]
\begin{tabular}{|c|c|c|}
\multicolumn{3}{c}{Table 4 } \\
\hline
$M_0$ (MeV) & $\Gamma_0$ (MeV) & References \\
\hline\hline
$1210.9 \pm 1.4$ & $106.5 \pm 3.5$ & [10] \\
\hline
$1210.30 \pm 0.36$ & $108.0 \pm 0.52$ & [11] \\
\hline
$1210.75 \pm 0.60$ & $105.6 \pm  1.2$ & [12] \\
\hline
$1212.60 \pm 0.52$ & $103.95 \pm 0.88$ & our results case C \\
\hline
$1214.00 \pm 0.53$ & $102.98 \pm 0.85$ & our results case (a)\\
\hline
\end{tabular}
\end{table}
\newpage

\

\begin{table}[h]
\begin{tabular}{|c|c|c|c|c|}
\multicolumn{5}{c}{Table 5} \\
\hline
  & $M_0-M_{++}$ & $\Gamma_0-\Gamma_{++}$ & $\wt{M}_0-\wt{M}_{++}$ &
$\wt{\Gamma}_0 -\wt{\Gamma}_{++}$ \\
\hline\hline
Our results I &$0.40 \pm 0.57$ & $6.89 \pm 0.95$ & $2.25 \pm 0.68$ &
$8.45\pm 1.11$ \\
Our results II &$0.70 \pm 0.70$ & $6.73 \pm 1.09$ &$2.47 \pm 0.80$ &
$8.51 \pm 1.26$ \\
\hline
Pedroni {\em et al} [5]  & --  & -- & $1.4 \pm 0.3$ & $8.6 \pm 1.0$ \\
\hline
Koch {\em et al} [13]  & --  & -- & $2.7 \pm 0.6$ & $2.0 \pm 1.8$ \\
\hline
Zidell {\em et al} [11] & $-0.40 \pm 0.39$ & $8.79 \pm 0.57$ & $1.9 \pm
0.4$ & $8.1 \pm 0.5$ \\
\hline
Vasan {\em et al} [12] & $1.15 \pm 0.78$ & $4.8 \pm 1.6$ & -- & -- \\
\hline
\end{tabular}
\end{table}
\newpage

\

\begin{table}[h]
\begin{tabular}{|c|c|c|}
\multicolumn{3}{c}{Table 6 } \\
\hline
 & Case (a) & Case (b) \\
\hline\hline
$M_{++}$ (MeV)& $1213.30 \pm 0.23$ & $1213.70 \pm 0.26$ \\
\hline
$\Gamma_{++}$ (MeV) & $96.17 \pm 0.34$ & $96.60 \pm 0.43$ \\
\hline
$a_{++}$ & $0.7175 \pm 0.0189$ &$0.7725 \pm 0.0285$ \\
\hline
$x_{++} (\wt{M}^2)$ & $- 0.3868 \pm 0.0014$ & $- 0.3794 \pm 0.0017$ \\
\hline
$\wt{M}_{++}$ (MeV) & $1231.89 \pm 0.25$ & $1232.02 \pm 0.29$ \\
\hline
$\wt{\Gamma}_{++}$ (MeV) & $108.04 \pm 0.39$ & $107.97 \pm 0.50$ \\
\hline
\end{tabular}
\end{table}

\begin{table}[h]
\begin{tabular}{|c|c|c|}
\multicolumn{3}{c}{Table 7 } \\
\hline
 & Case (a) & Case (b) \\
\hline\hline
$M_0$ (MeV)& $1214.00 \pm 0.53$ & $1214.50 \pm 0.71$ \\
\hline
$\Gamma_0$ (MeV) & $102.98 \pm 0.85$ & $103.11 \pm 0.93$  \\
\hline
$a_0$ & $0.8516 \pm 0.0623$ & $0.9056 \pm 0.0826$ \\
\hline
$x_0 (\wt{M}^2)$ & $- 0.4033 \pm  0.0033$ & $- 0.4012 \pm 0.0036$ \\
\hline
$\wt{M}_0$ (MeV) & $1234.76 \pm 0.58$ & $1235.17 \pm 0.78$ \\
\hline
$\wt{\Gamma}_0$ (MeV) & $117.88 \pm 1.01$ & $117.76 \pm 1.10$ \\
\hline
$\epsilon$ ($\times$ 10$^{-2}$) & $2.3 \pm 0.3$ & $2.5 \pm 0.4$ \\
\hline
\end{tabular}
\end{table}

\end{document}